\documentclass[a4paper]{article}
\usepackage{epsfig}

\newcommand{\tensor}{\otimes}

\newcommand{\unit}{\mathbf{1}}
\newcommand{\re}{\mathbbm{R} \mbox{e }}

\usepackage{amsfonts}
\usepackage{bbm}

\begin{document}


\title{A Bayesian Account of Quantum Histories}\author{Thomas Marlow\thanks{email: pmxtm@nottingham.ac.uk}\\ \emph{School of Mathematical Sciences, University of Nottingham,}\\
\emph{UK, NG7 2RD}}

\maketitle

\begin{abstract}
We investigate whether quantum history theories can be consistent with Bayesian reasoning and whether such an analysis helps clarify the interpretation of such theories.  First, we summarise and extend recent work categorising two different approaches to formalising multi-time measurements in quantum theory.  The standard approach consists of describing an ordered series of measurements in terms of history propositions with non-additive `probabilities'.  The non-standard approach consists of defining multi-time measurements to consist of sets of exclusive and exhaustive history propositions and recovering the single-time exclusivity of results when discussing single-time history propositions.  We analyse whether such history propositions can be consistent with Bayes' rule.  We show that certain class of histories are given a natural Bayesian interpretation, namely the linearly positive histories originally introduced by Goldstein and Page.  Thus we argue that this gives a certain amount of interpretational clarity to the non-standard approach.  We also attempt a justification of our analysis using Cox's axioms of probability theory.
\end{abstract}



\section*{Outline}

The basic premise of this paper is rather simple.  We propose to apply Bayesian probability rules to quantum histories theory and see if we get any form of consistency.  The are a few reasons why this is a pedagogically useful tack to take.  Firstly, Bayesian probability is pedagogically useful in its own right as it provides a framework for thinking about probabilities that is rather natural in a human sense---it accommodates, in different situations, all uses of the term `probability' including probabilistic inference and relative frequencies \cite{JaynesBOOK}.  Secondly, quantum history theories are specifically designed with the idea of applying such probabilities to closed systems, without necessarily discussing observers and their experiments; thus it is natural to interpret such probabilities in a Bayesian manner rather than necessarily discussing the relative frequencies of experiments.  In fact Bayesian probability can accommodate almost all notions of relative frequency presently used in the literature \cite{JaynesBOOK}, whereas theories of relative frequency have to be designed for the problem at hand.  Thirdly, even when discussing quantum histories instrumentally such a Bayesian interpretation might help to clarify the interpretation of Standard Quantum Theory (SQT) \cite{Mana04}.  It will turn out that we can apply a very natural Bayesian interpretation to a certain class of quantum histories.  So, although it may seem na\"{\i}ve at first we will get out something quite profound, a natural interpretation of the probabilities of certain history propositions.  In fact, as we will show, we can justify our analysis using Cox's axioms of probability theory \cite{CoxBOOK} to show that the standard notions of probability in the consistent histories programme aren't necessarily `good' notions of probability, but there are alternative notions.

Before we get stuck into our Bayesian analysis, let us briefly discuss why the foundational interpretation of probability really matters when interpreting such theories.  Then we will introduce quantum history theories and try and analyse what consistency we can get through Bayesian reasoning.

\section*{Quantum Probabilities}

In Standard Quantum Theory (SQT) one usually invokes multi-time measurements as a succession of single-time measurements. If we use the von Neumann measurement formalism then the possible exclusive propositions at each time are represented by the projection operators associated with the eigenstates of a Hermitian operator.  Thus a non-relativistic history is represented as a succession of such projection operators each labelled by a time.  The probability of each such history is, in the standard formalism, given by the probability trace formula.  For example, for a three-time succession ($t_1 < t_2 < t_3$) of von Neumann measurements $\{\hat{A}(t_1), \hat{B}(t_2), \hat{C}(t_3)\}$ given an initial state $\rho$, the probability of a history $\{\hat{a}_i(t_1), \hat{b}_j(t_2), \hat{c}_k(t_3)\}$ is given by:

\begin{equation}
p(a_i, b_j, c_k \vert \rho) = \mbox{tr}(\hat{c}_k(t_3) \hat{b}_j(t_2)  \hat{a}_i(t_1) \rho \hat{a}_i(t_1) \hat{b}_j(t_2)).
\label{Trace}
\end{equation}

In the above equation the results of each measurement are conveniently represented by Heisenberg picture projection operators.  For example, the results of the von Neumann measurement $\hat{A}$ at time $t_1$ are represented by the set of Heisenberg picture projection operators $\hat{a}_i(t_1) = U^\dagger(t_1-t_0) \hat{a}_i U(t_1-t_0)$ where $\hat{a}_i$ are the relevant Schr\"odinger picture operators and $t_0$ is the fiducial time.  Similarly for the other von Neumann measurements $\hat{B}$ and $\hat{C}$.  This is the standard way that multi-time measurements are invoked in non-relativistic SQT.

Classically, the additivity of propositions in Bayesian probability theory is a contextual property of propositions.  This can be seen in the pedagogical example given recently by Mana \cite{Mana04}.  Take an urn that contains some red balls and some wooden balls; the urn is shaken and an observer takes out a ball.  We can ask for the probabilities of the following two propositions: ``the ball is red'' and ``the ball is wooden''.  Only if it is the case that the balls cannot be both wooden and red then these two propositions are exclusive.  Thus, we can see that propositions are not \emph{inherently} exclusive.  Rather, classically at least, exclusivity is a contextual property of propositions as there are ways that these two propositions could be non-exclusive (say, some balls are both red and wooden). Therefore, since there is no mention of contexts in the standard analysis, one is not \emph{necessarily} discussing exclusive propositions when discussing the possible history propositions that arise through an ordered succession of von Neumann measurements.  Or, one might, ambiguously, be implicitly invoking many possible contexts that need to be formally differentiated.

The \emph{only} way, classically, we have to define whether two propositions $A$ and $B$ are exclusive is to equate exclusivity with the additivity of their probabilities:

\begin{equation}
p(A \cup B \vert I) = p(A \vert I) + p(B \vert I)
\label{ADD}
\end{equation}

\noindent such that $A \cap B = \emptyset$ where `$\emptyset$' is the null proposition that is always false in standard Boolean logic---when $A \cap B = \emptyset$ we say that $A$ and $B$ are disjoint.  So, if two propositions are both additive and disjoint we will simply call them `exclusive' with respect to context $I$.  If this is not satisfied then propositions $A$ and $B$ are called `not-exclusive' with respect to context $I$.   Classical probability theory and SQT therefore differ by how they treat `not-exclusive' propositions.  Note that we use the term `exclusive', throughout this paper, in a pedagogically distinct way to how it is normally used in the quantum histories literature---where exclusive is synonymous with disjoint.  We wish to differentiate `disjoint' and `exclusive' propositions because, classically, exclusive propositions must always be additive so we wish to reserve the word `exclusive' only for contextually additive propositions.  This means that when we generalise we keep the standard notion of exclusivity and are forced to name any other tentative notion something else so as to avoid confusion.  In standard Bayesian probability theory exclusive and disjoint are considered equivalent notions (the former being about probabilities and the latter about the propositions themselves) but when we get into problems with non-additivity we should differentiate these notions.  Obviously, single-time von Neumann measurements consist of sets of `exclusive' propositions (both additive and disjoint).  When we wish to differentiate our introduced notion of conventional probabilistic exclusivity from other presumed notions of exclusivity we will do so explicitly, otherwise we will simply use the term exclusive in the standard contextual probabilistic way we have introduced above.  When we come to discuss quantum history theories it will turn out that exclusive propositions are also disjoint, but disjoint propositions aren't necessarily exclusive (since exclusivity is taken to be a contextual probabilistic property of propositions whereas disjointness is something that is defined on the proposition algebra).

Here we use standard Bayesian notation such that all probabilities are defined with respect to a specific context $I$ for exactly the reason noted above: propositions $A$ and $B$ are only well-defined in a given context exactly because their meaning is contextual\footnote{We use the term `context' in the colloquial manner used by Bayesian theorists rather than in the technical sense of the Kochen-Specker theorem.}.  By invoking contexts explicitly we hope to clarify the meaning of such statements.  Since the probabilities given by (\ref{Trace}) are not necessarily additive then we are, in the standard interpretation, having to invoke a different kind of exclusivity to that invoked when requiring both (\ref{ADD}) and disjointness.  The standard von Neumann interpretation of exclusivity comes about because each single-time measurement consists of explicitly exclusive propositions and it is rather natural (although we argue that it is perhaps dubious) to \emph{presume} that successions of these single-time measurements give well-defined exclusive history propositions---even though the non-additivity of the probabilities of such history propositions suggests that they are not probabilistically `exclusive' as we have defined above.

So, as Anastopoulos has argued \cite{Anast04}, there is a dichotomy for multi-time measurements that we must account for.  We have a choice between the two following paradigms:

\begin{enumerate}

\item Postulate single-time exclusivity of results in the standard manner and presume some na\"{\i}ve kind of exclusivity for multi-time propositions that arise from a series of single-time measurements.  This is the standard interpretation of von Neumann measurements.
\label{1}

\item Postulate the exclusivity of some history propositions, using the standard notion of exclusivity of probability theory as we have defined above, and get single-time exclusivity of results as a corollary by discussing single-time history propositions.
\label{2}

\end{enumerate}

In what follows we shall refer to these two interpretations as \ref{1} and \ref{2} respectively.  Anastopoulos argues, in \cite{Anast04}, that neither interpretation of multi-time measurements have yet been convincingly promoted.  We are, of course, used to interpretation \ref{1} and not used to interpretation \ref{2}.  If we use interpretation \ref{1} then, as Anastopoulos \cite{Anast04} shows, we are forced to admit a dependency of the probabilities (treated as relative frequencies) on the resolution of the apparatus we use, exactly because such `probabilities' are non-additive and thus aren't exclusive in the conventional sense (nor are they not-exclusive in the conventional sense).  So, if we use finer-grained projection operators we get different probabilities out for given sample sets.  It is interesting to investigate interpretation \ref{2} simply because it is not usually considered and cannot be rejected \emph{a priori}.

It is the conflict implicit in the noted dichotomy which makes us so uncomfortable with multi-time measurements.  In interpretation \ref{1}, \emph{any} ordered set of measurements, presuming that the relevant apparatus can be made, is well-defined---this suggests an amazing amount of freedom that nature gives to experimental physicists.

\section*{A Pedagogical Account of Consistent Histories}

There does exist a quantum formalism that implicitly uses interpretation \ref{2}; namely the Consistent Histories (CH) programme \cite{Grif84, Omnes88, GH90, Isham94}. Rarely, however, is interpretation \ref{2} explicitly used by consistent historians.  Rather, CH is usually invoked in a non-instrumental fashion (some exceptions to this trend are \cite{Hartle04} and \cite{GriffithsBOOK}).  Interpretation \ref{2} is also in opposition to the general claim by some consistent historians that CH solves the measurement problem.  Interpretation \ref{2} is a way to \emph{re-define} measurement rather than solve the measurement problem \emph{per se}.

Let us give a brief introduction to the CH programme; the basic setup of CH is as follows.  One defines a set of homogeneous history propositions; following \cite{Isham94}, each homogeneous history proposition $\alpha$ consists of an ordered tensor product of time-labelled projection operators just like in SQT---for example:

\begin{equation}
\alpha := \hat{\alpha}_{t_n}(t_n) \tensor \hat{\alpha}_{t_{n-1}}(t_{n-1}) \tensor ...\hat{\alpha}_{t_2}(t_2) \tensor \hat{\alpha}_{t_1}(t_1)
\end{equation}

\noindent where each $\hat{\alpha}$ is a standard single-time projection operator.  Here we use the Heisenberg picture. The ordered set of times over which an homogeneous history is defined is called its temporal support.  We can then naturally define the class operator \cite{Isham94} for such a history to be:

\begin{equation}
C_{\alpha} := \hat{\alpha}_{t_n}(t_n) \hat{\alpha}_{t_{n-1}}(t_{n-1}) ...\hat{\alpha}_{t_2}(t_2) \hat{\alpha}_{t_1}(t_1)
\end{equation}

\noindent and the probability formula (\ref{Trace}) becomes:

\begin{equation}
p(\alpha \vert I) = \mbox{tr}(C_{\alpha} \rho C^{\dagger}_{\alpha}).
\end{equation}

It is natural to extend the definition of history propositions to include inhomogeneous history propositions \cite{Isham94}.  Inhomogeneous history propositions are defined by combining homogeneous history propositions in novel, but rather natural, ways.  We will not repeat such arguments here (see Isham's original work \cite{Isham94}) because it is sufficient simply to note the following.  One can define `or' and `not' operations for homogeneous history propositions in a rather natural manner; such operations are denoted `$\vee$' and `$\neg$' respectively.  These operations are not the standard notions of `or' and `not' in Boolean logic, but are defined naturally on the history algebra.  The standard `and' operation `$\wedge$' takes homogeneous history propositions into homogeneous history propositions and behaves exactly like the Boolean `and' operation should. We can also naturally define a notion of disjointness; we denote such a relation `$\perp$'.  Note that we have explicitly been calling these histories `propositions'; this is because, in analogy with Bayesian probability theory, we are going to treat them as propositions in the standard sense to see if we get any consistency via Bayesian reasoning.

When two homogeneous history propositions $\alpha$ and $\beta$ are disjoint (such that they have the same temporal support) then the class operator for the history $\alpha \vee \beta$ is simply:

\begin{equation}
C_{\alpha \vee \beta} = C_{\alpha} + C_{\beta}.
\end{equation}

We define two history propositions to be `exclusive' if their probabilities are additive under this `$\vee$' operation and such that, in the \emph{same} context, the probability of both being the case is zero.  A sufficient condition for two disjoint history propositions to have additive probabilities, and thus be exclusive propositions, is defined using what is called the decoherence functional $d$.  For SQT the decoherence functional acting on two homogeneous history propositions $\alpha$ and $\beta$ is defined as follows:

\begin{equation}
d_{\rho, H}(\alpha, \beta) := \mbox{tr}(C_{\alpha} \rho C_{\beta}^\dagger).
\label{dFUNC}
\end{equation}

There is an ambiguity in how we have defined homogeneous history propositions because we have used the Heisenberg picture in their definition, but obviously one could use Schr\"{o}dinger picture projection operators and absorb all the dynamics into the definition of the decoherence functional.  So, the subscripts $\rho$ and $H$ refer to such a dependence of the decoherence functional on the initial state and the Hamiltonian.  We will drop these subscripts from now on and such dependence is kept implicit.  One can consider the initial state and dynamics constant throughtout the following discussion.  Obviously, $d(\alpha, \alpha)$ has the same form as (\ref{Trace}).  If we take two homogeneous history propositions $\alpha^i$ and $\alpha^j$ then their respective probabilities ($d(\alpha^i, \alpha^i)$ and $d(\alpha^j, \alpha^j)$ respectively) are additive if $d(\alpha^i, \alpha^j) = 0$---if this is the case then we will call these two history propositions `exclusive' with respect to a context $I$.  We call the context `$I$' simply to give it a name and invoke a context explicitly---we reserve the right to change its name, or use a different context, later.  $I$ obviously must specify $\rho$ and $H$, but it may also specify further information at present left unspecified.  A set of such history propositions $\{\alpha^i: i=1,2,...,N\}$ is called `$d$-consistent' \cite{Isham94} when all such propositions are mutually `exclusive' and exhaustive with respect to context $I$.  Single-time SQT is recovered by noting that von Neumann single-time measurements are $d$-consistent sets of single-time history propositions.  For two disjoint histories we have that $d(\alpha^i \wedge \alpha^j, \alpha^i \wedge \alpha^j) = 0$.  A set of disjoint histories that form a partition of unity such that $\sum_i d(\alpha^i, \alpha^i) = 1$ is simply called a `complete' set.

If we use interpretation \ref{2} then it is clear that we could equate multi-time measurements with $d$-consistent sets.  However, rather than call a $d$-consistent set a measurement (which might get quite confusing when discussing the distinction between interpretations \ref{1} and \ref{2}) we will call a $d$-consistent set a `null-counterfactual'.  A null-counterfactual consists of an exclusive and exhaustive set of propositions.  A counterfactual statement is a statement about what would have happened in a different context; a null-counterfactual statement is simply the trivial statement about what would happen if the same context was invoked.  Obviously in quantum theory a null-counterfactual can have many different exclusive results because of its probabilistic nature \cite{Grif98}.  So a null-counterfactual is almost like a definition of `context', but we do not wish to use the term `context' because of its more technical use in SQT and because, in what follows, we use the term in the more colloquial Bayesian manner.  If such a von Neumann measurement is repeated using an identical setup then one of the possible propositions is, exclusively, the case; so, a standard single von Neumann measurement is a null-counterfactual.  The same is considered true for null-counterfactuals consisting of more general history propositions.  A series of von Neumann measurements does not necessarily define a null-counterfactual.

Null-counterfactuals can also include inhomogeneous history propositions.  Some history propositions cannot be defined in any $d$-consistent set; such history propositions are to be called non-$d$-realisable.  A necessary and sufficient condition for a history proposition $\alpha^i$ to be $d$-realisable is thus:

\begin{equation}
d(\alpha^i, \alpha^i) + d(\neg \alpha^i, \neg \alpha^i) = 1.
\end{equation}

Although it is not yet clear which interpretation which out of \ref{1} or \ref{2} is physically correct, it is pedagogically interesting to investigate interpretation \ref{2} because it is not usually considered and cannot be rejected \emph{a priori} \cite{Anast04}.  Adopting \ref{2} is tempting because of its clear and unambiguous definition of exclusivity and null-counterfactual statements.  In interpretation \ref{1} one might run an experiment and a certain history proposition is realised;  one may then ask a null-counterfactual question: ``what history propositions could be realised if you repeated the experiment in \emph{exactly the same manner}?'' and you are forced to \emph{presume} that any distinct history proposition that is realised upon a second run is exclusive to the one you first received even though it is not probabilistically exclusive in the standard sense.  Using null-counterfactuals in interpretation \ref{2} one bypasses this problem (as such a definition of exclusivity is uncontroversial).  Just as we define a single-time measurement to be some kind of context in which an exclusive set of single-time propositions can be realised, so it seems we might wish to define a multi-time measurement to be related to contexts in which an exclusive set of history propositions can be realised.

Just as von Neumann measurements can be convexly mixed we might assume that more general null-counterfactuals can be mixed.  If we use interpretation \ref{1} then it is clear that a succession of von Neumann measurements might not be defined by a $d$-consistent set of homogeneous history propositions, but each homogeneous history proposition might be $d$-realisable.  In such a case then perhaps, one might na\"{\i}vely think, we can define such a multi-time measurement using interpretation \ref{2} by mixing null-counterfactuals.

Note that in the CH interpretation of quantum systems the \emph{values} of probabilities of history propositions are independent of the $d$-consistent set they are taken to be part of.  This is rather analogous to the Gleason non-contextuality of single-time SQT \cite{ILS94}.  It is exactly this type of non-contextuality that has, in the history of SQT, confused the distinction between interpretations \ref{1} and \ref{2}.  This is because the independence of the values of probabilities upon contexts doesn't necessarily mean that we can disregard the contextual element of their very definition.  A tautology: probabilities with the same values are not necessarily the same probabilities---they might need to be distinguished. No two equals are the same.

One can easily imagine a situation where an experimenter mixes a set of von Neumann measurements such that she chooses each such measurement with a given weight.  In such a case it doesn't matter that the propositions realised within different measurements are not considered exclusive when taken together, they become exclusive only by the application of the mixing process.  Similarly one might be able to give a good definition to a mixture of null-counterfactuals.

So, if we take a succession of von Neumann measurements and label the possible history propositions as $\{\alpha^i : i=1, 2..., N\}$ then this set need not be $d$-consistent but if all the $\alpha^i$ are $d$-realisable then the sets $\{\alpha^i, \neg \alpha^i\}$ will be $d$-consistent for each $i=1,2,...,N$.  Let us denote $p(\alpha^i \vert I) = d(\alpha^i, \alpha^i)$ in order to emphasise the probabilistic interpretation of the decoherence functional.  If we mix these null-counterfactuals we must assign weights $w_i$ to each such $d$-consistent set $\{\alpha^i, \neg \alpha^i\}$, just as we would if we were mixing a set of von Neumann measurements.  Presuming that the context $I$ is the same for each element of the mixture (we reserve the right to change this assumption later but it is easy to assume that the proposition $\alpha^i \vee \neg \alpha^i$ is equivalent to the proposition $\alpha^j \vee \neg \alpha^j$ since they are normally considered equivalent tautologies, our mixture $M$ being a weighted set of different tautologies in this case) then the probability for any history $\alpha^i$ to be received given such a mixed set $M$ is given by the following:

\begin{equation}
p(\alpha^i \vert M) = w_i p(\alpha^i \vert I) + \sum_{j \neq i}^{N} w_j p(\alpha^i \wedge \alpha^j \vert I) + \sum_{j \neq i}^{N} w_j p(\alpha^i \wedge \neg \alpha^j \vert I).
\label{yourmum}
\end{equation}

If we equate $p(\alpha^i \wedge \alpha^j \vert I)$ with $d(\alpha^i \wedge \alpha^j, \alpha^i \wedge \alpha^j)$ then we must note that $d(\alpha^i \wedge \alpha^j, \alpha^i \wedge \alpha^j)$ equals zero for all disjoint homogeneous history propositions $\alpha^i$ and $\alpha^j$ which are defined over the same temporal support. This is because, for all such history propositions, $\alpha^i \wedge \alpha^j \equiv \bf{0}$, where $\bf{0}$ is used to denote the null history proposition.  Thus for all homogeneous history propositions so defined the first summation in (\ref{yourmum}) is equal to zero.

The simplest case is when all the $d$-consistent sets $\{\alpha^i, \neg \alpha^i\}$ are \emph{a priori} equally likely; so lets try this and see what happens in the case where $w_i = \frac{1}{N}$ for all $i$---each $d$-consistent set will be the corresponding null-counterfactual with \emph{a priori} weight $\frac{1}{N}$.  Note that we don't yet call such weights `probabilities'.  In such a case we get:

\begin{equation}
p(\alpha^i \vert M) = \frac{1}{N}(p(\alpha^i \vert I) + \sum_{j \neq i}^{N} p(\alpha^i \wedge \neg \alpha^j \vert I)).
\end{equation}

Now we must ask what form $p(\alpha^i \wedge \neg \alpha^j \vert I)$ should take in terms of the decoherence functional.  It is clear that, by intuition, the history proposition $\alpha^i \wedge \neg \alpha^j$ is equivalent to $\alpha^i$.  The proposition that the history proposition $\alpha^i$ is the case \emph{and} the proposition that $\alpha^j$ isn't the case is just equivalent to the proposition that the history proposition $\alpha^i$ is the case. This can be shown explicitly in the History Projection Operator (HPO) form of CH \cite{Isham94}.  In the HPO formalism homogeneous history propositions are represented by tensor products of the relevant single-time propositions.  So for two two-time history propositions we have that $\alpha^i = \hat{\alpha}^{i}_{t_1} \tensor \hat{\alpha}^{i}_{t_2}$ and $\alpha^j = \hat{\alpha}^{j}_{t_1} \tensor \hat{\alpha}^{j}_{t_2}$.  If we assume that these two history propositions are defined such that $\hat{\alpha}^{i}_{t_1} \perp \hat{\alpha}^{j}_{t_1}$ and $\hat{\alpha}^{i}_{t_2} \perp \hat{\alpha}^{j}_{t_2}$ then the proof that $\alpha_i \wedge \neg \alpha_j = \alpha_i$ goes as follows:

\begin{eqnarray}
\hat{\alpha}^{i}_{t_1} \tensor \hat{\alpha}^{i}_{t_2} \wedge \neg (\hat{\alpha}^{j}_{t_1} \tensor \hat{\alpha}^{j}_{t_2})
&:=& \hat{\alpha}^{i}_{t_1} \wedge \neg \hat{\alpha}^{j}_{t_1} \tensor \hat{\alpha}^{i}_{t_2} \wedge \neg \hat{\alpha}^{j}_{t_2} \nonumber \\
& & + \ \hat{\alpha}^{i}_{t_1} \wedge \neg \hat{\alpha}^{j}_{t_1} \tensor \hat{\alpha}^{i}_{t_2} \wedge \hat{\alpha}^{j}_{t_2} \nonumber \\
& & + \ \hat{\alpha}^{i}_{t_1} \wedge \hat{\alpha}^{j}_{t_1} \tensor \hat{\alpha}^{i}_{t_2} \wedge \neg \hat{\alpha}^{j}_{t_2} \label{And}\\
&=& \hat{\alpha}^{i}_{t_1} \tensor \hat{\alpha}^{j}_{t_2} + \hat{\alpha}^{i}_{t_1} \tensor \hat{\bf{0}} + \hat{\bf{0}} \tensor \hat{\alpha}^{j}_{t_2} \label{Nullhistory propositions}\\
&=& \hat{\alpha}^{i}_{t_1} \tensor \hat{\alpha}^{i}_{t_2} \label{Answer}.
\end{eqnarray}

Eq.(\ref{And}) represents the intuitive logical result that the history propostion $\alpha^i \wedge \neg \alpha^j$ can be true in three different ways, namely if any one of the three history propositions on the RHS of Eq.(\ref{And}) is true.  To get from Eq.(\ref{Nullhistory propositions}) to Eq.(\ref{Answer}) we simply note that all history propositions which have a null result at any given time are deemed equivalent to the null history $\bf{0} = \hat{\bf{0}} \tensor \hat{\bf{0}}$ \cite{Isham94}.

Thus, for homogeneous history propositions that are defined using exclusive and exhaustive single-time propositions, it is always the case that $\alpha^i \wedge \neg \alpha^j = \alpha^i$ and thus that:

\begin{equation}
d(\alpha^i \wedge \neg \alpha^j, \alpha^i \wedge \neg \alpha^j) = d(\alpha^i, \alpha^i).
\end{equation}

Thus it seems natural to equate $p(\alpha^i \wedge \neg \alpha^j \vert I)$ with $p(\alpha^i \vert I) = d(\alpha^i, \alpha^i)$ in this case.  This gives us that the probability that history proposition $\alpha^i$ is the case, given that context $M$ is an equally weighted mixture of null-counterfactuals, is given by the rather trivial result:

\begin{eqnarray}
p(\alpha^i \vert M) &=& \frac{1}{N}(p(\alpha^i \vert I) + p(\alpha^i \vert I) \sum_{j \neq i}^{N} 1) \\ &=& \frac{1}{N}(p(\alpha^i \vert I) + (N-1)p(\alpha^i \vert I)) \\ &=& p(\alpha^i \vert I) = d(\alpha^i, \alpha^i).
\label{MIXTURE}
\end{eqnarray}

Note that this result does not depend upon the fact that we chose to use equal weights.  Any set of positive weights $w_i$, such that $\sum_i w_i = 1$, would work.  This result should be taken with a pinch of salt, lots of implicit assumptions have been made in order to reach (\ref{MIXTURE})---we will investigate it less na\"{\i}vely in the next section once we have introduced a Bayesian account of such propositions.  So, it is clear that some ordered sets of single-time von Neumann measurements might equally well be interpreted as a mixed set of null-counterfactuals---but, of course, not all ordered sets of von Neumann measurements could be interpreted in such a manner.

Non-$d$-realisable propositions are propositions that can never be $d$-realised with respect to some other history propositions in context $I$.  So far we have not discussed the strict meaning of the context $I$; we have simply kept the context within the notation because of the tentative contextual meaning we apply to propositions.  In \cite{Mana04} it was shown that doing such a thing can help clarify the meaning of probabilistic statements in SQT; we simply adopt the same principle here for quantum history theories. So, for the moment, one is asked just to accept the name `$I$' for the the context whatever that context is taken to mean.  We will, in part, rectify this gnomic situation later.

So, strictly speaking, the above discussion is rather na\"{\i}ve and we have yet to check that the reasoning we have used is all consistent and unambiguous.  For example, it is not clear that the context $I$ is well-defined globally throughout the mixing process.  Or whether the mixing process is itself well-defined---especially since the weights are totally arbitrary.  We shall examine this in the following sections.  We will show that by using Bayesian reasoning such concepts do become consistent and less ambiguous.

\section*{Bayesian Histories}

Bayes' rule is a rule that relates \emph{a priori} probability statements to \emph{a posteriori} probability statements.  Say we have two propositions $A$ and $B$ and a general context $D$ which refers to the general setup of the problem (and remains constant through the analysis) then Bayes' rule is as follows:

\begin{equation}
p(A \vert B D) = \frac{p(B \vert A D)p(A \vert D)}{p(B \vert D)}.
\label{BAYES}
\end{equation}

\noindent Bayes' rule is derived from the following rule:

\begin{equation}
p(A \cap B \vert D) = p(A \vert B D) p(B \vert D) = p(B \vert A D)p(A \vert D).
\label{AND}
\end{equation}

We can try and use Bayes' rules (or equivalently (\ref{AND})) to analyse the reasoning we used above.  If we take all history propositions then one might be tempted to try and apply Bayes' rule to them and see if we get any form of consistency.  So, let us apply Bayes' rule in the following na\"{\i}ve way, simply using the history algebra `$\wedge$' instead of the standard Boolean `$\cap$' (we will justify the step using Cox's axioms of probability later):

\begin{eqnarray}
p(\alpha^i \wedge \neg \alpha^j \vert I) &=& p(\alpha^i \vert \neg \alpha^j I) p(\neg \alpha^j \vert I) \\ &=& p(\neg \alpha^j \vert \alpha^i I) p(\alpha^i \vert I).
\end{eqnarray}

By intuition one might like to assign that $p(\neg \alpha^j \vert \alpha^i I) := 1$ because if the proposition $\alpha^i$ is true then obviously the proposition $\neg \alpha^j$ is true. This then gives us that:

\begin{equation}
p(\alpha^i \wedge \neg \alpha^j \vert I) = p(\alpha^i \vert I).
\label{gay}
\end{equation}

The above analysis is consistent as long as Bayes' rule is valid for such history propositions, so we need to work out if Bayes' rule is a valid way to manipulate the probabilities of history propositions.  For this to be the case then all the above probabilities must be well-defined.  We will justify our na\"{\i}ve application of Bayes' rule later, but for now let us continue along with this na\"{\i}ve analysis for a moment and see how Bayes' rule (or equivalently rule (\ref{AND})) apply to the other probability assignments we might like to make.

\begin{eqnarray}
p(\alpha^i \wedge \alpha^j \vert I) &=& p(\alpha^i \vert \alpha^j I) p(\alpha^j \vert I) = 0 \label{ONE} \\ &=& p(\alpha^j \vert \alpha^i I) p(\alpha^j \vert I) = 0. \label{TWO}
\end{eqnarray}

The statement (\ref{ONE}) is intuitively the case for disjoint history propositions since if $\alpha^j$ is the case then $\alpha^i$ isn't the case, and similarly for the second decomposition (\ref{TWO}). Note that this doesn't presume that these two propositions are probabilistically exclusive, only that given one we never infer the other.

\begin{eqnarray}
p(\neg \alpha^i \wedge \alpha^j \vert I) &=& p(\neg \alpha^i \vert \alpha^j I) p(\alpha^j \vert I) \\ &=& p(\alpha^j \vert I). \label{gay2}
\end{eqnarray}

To get to (\ref{gay2}) we use exactly the same reasoning we used to get (\ref{gay}).  And so we come to ask how we interpret $p(\neg \alpha^i \wedge \neg \alpha^j \vert I)$.  One way to look at this probability is to decompose it as follows:

\begin{eqnarray}
p(\neg \alpha^i \wedge \neg \alpha^j \vert I) &=& p(\neg \alpha^j \vert \neg \alpha^i I) p(\neg \alpha^i \vert I) \label{DONE} \\ &=& (1 - p(\alpha^j \vert \neg \alpha^i I))p(\neg \alpha^i \vert I) \label{29}\\ &=& (1 - \frac{p(\alpha^j \vert I)}{p(\neg \alpha^i \vert I)}) p(\neg \alpha^i \vert I) \\ &=& p(\neg \alpha^i \vert I) - p(\alpha^j \vert I).
\end{eqnarray}

\noindent But, of course, instead of using the decomposition (\ref{DONE}) one could have used:

\begin{eqnarray}
p(\neg \alpha^i \wedge \neg \alpha^j \vert I) &=& p(\neg \alpha^i \vert \neg \alpha^j I) p(\neg \alpha^j \vert I) \label{DTWO}\\ &=& (1 - p(\alpha^i \vert \neg \alpha^j I))p(\neg \alpha^j \vert I) \label{33}\\ &=& (1 - \frac{p(\alpha^i \vert I)}{p(\neg \alpha^j \vert I)}) p(\neg \alpha^j \vert I) \\ &=& p(\neg \alpha^j \vert I) - p(\alpha^i \vert I).
\end{eqnarray}

In order for the two ways of decomposing $p(\neg \alpha^i \wedge \neg \alpha^j \vert I)$ to be consistent we require that $p(\neg \alpha^j \vert I) - p(\alpha^i \vert I) = p(\neg \alpha^i \vert I) - p(\alpha^j \vert I)$.  A necessary and sufficient condition for the history propositions to satisfy this requirement is that:

\begin{equation}
p(\alpha^i \vert I) + p(\neg \alpha^i \vert I) = K \mbox{ for all } i,
\label{QUASI}
\end{equation}

\noindent where $K$ is a positive constant.  We call condition (\ref{QUASI}) quasi-realisability.  When $K=1$ we call the probabilities realisable.  Note that in assuming steps (\ref{29}) and (\ref{33}) are valid we must presume that the probabilities are realisable in an \emph{a posteriori} sense in that $p(\alpha^i \vert \neg \alpha^j I) + p(\neg \alpha^i \vert \neg \alpha^j I) = 1$ for all $i$.

A set $\{\alpha^i : i=1,2,...,N\}$ that does \emph{not} satisfy (\ref{QUASI}) does not give equal decompositions (\ref{DONE}) and (\ref{DTWO}).  Thus, in interpretation \ref{1} any complete set of homogeneous history propositions makes sense, but if we require consistency with Bayesian probability theory then we must at least discuss sets of history propositions which satisfy the stricter condition (\ref{QUASI}).

So, if we identify that $p(\alpha^i \vert I) = d(\alpha^i, \alpha^i)$ and that $p(\neg \alpha^i \vert I) = d(\neg \alpha^i, \neg \alpha^i)$, then a sufficient condition for all the above to be consistent by both decompositions (\ref{DONE}) and (\ref{DTWO}) is that everything is $d$-realisable.  If the history propositions weren't $d$-realisable then there is no \emph{a priori} reason why decompositions (\ref{DONE}) and (\ref{DTWO}) should match.  However, are all these probabilities well-defined?  All the probabilities that we identify with decoherence probabilities are obviously well-defined in the sense that they are bounded between 0 and 1.  But we haven't yet identified whether the na\"{\i}ve conditional probabilities are all well-defined.

For example, if we make the identification that $p(\neg \alpha^j \vert \alpha^i I) = 1$ and then, using Bayes' rule, we derive:

\begin{equation}
p(\alpha^i \vert \neg \alpha^j I) = \frac{p(\alpha^i \vert I)}{p(\neg \alpha^j \vert I)} = \frac{d(\alpha^i, \alpha^i)}{d(\neg \alpha^j, \neg \alpha^j)}.
\end{equation}

In order for the above to be bounded by 0 and 1 we require that:

\begin{equation}
0 \leq \frac{d(\alpha^i, \alpha^i)}{d(\neg \alpha^j, \neg \alpha^j)} \leq 1.
\label{CONDITION}
\end{equation}

If $\alpha^i$ is more probable than $\neg \alpha^j$ in the context $I$ then the above condition will not be satisfied.  The next question we must ask is what types of history propositions do we require for this Bayesian analysis to be consistent?  In terms of the HPO form of CH \cite{Isham94} we define that the the homogeneous history propositions $\{\alpha^i : i = 1,2,...,N\}$ defined using exclusive sets of single-time propositions are all mutually disjoint: $\alpha^i \perp \alpha^j$ for all $i,j$ such that $i \neq j$.  In terms of the natural orthoalgebra of history propositions, this means that $\alpha^i \leq \neg \alpha^j$ for all $i,j$ such that $i \neq j$.  Thus if the decoherence functional preserves the partial order defined on the history proposition space then condition (\ref{CONDITION}) is satisfied.  Isham and Linden \cite{IL94} have argued that this need not be the case; there are examples where the following is not true:

\begin{equation}
\alpha \leq \beta \Rightarrow d(\alpha, \alpha) \leq d(\beta, \beta).
\label{PRESERVE}
\end{equation}

They give a specific example which disobeys (\ref{PRESERVE}).  The sum-over-paths formulation for SQT does obey (\ref{PRESERVE}) and it is thus not clear whether we should assume it in general history theories \cite{IL94}. In order to satisfy (\ref{CONDITION}), however, we \emph{must} use sets of history propositions such that the following is satisfied:

\begin{equation}
\alpha^i \leq \neg \alpha^j \Rightarrow d(\alpha^i, \alpha^i) \leq d(\neg \alpha^j, \neg \alpha^j) \mbox{ for all } i \neq j.
\label{PRESERVE2}
\end{equation}

Presuming $\{\alpha^i: i=1,2,...,N\}$ consists of $d$-realisable homogeneous history propositions that are defined using exclusive and exhaustive single-time propositions then this Bayesian analysis is consistent as long as (\ref{PRESERVE2}) is satisfied.  If this is the case then the probabilities $p(\alpha^i \vee \alpha^j \vert \alpha^k I)$ and $p(\alpha^i \vee \alpha^j \vert \neg \alpha^k I)$ are well-defined (in the sense of being bounded by 0 and 1) for all $i,j,k$.  This can be seen just by invoking the conditional probabilities invoked above.

\begin{eqnarray}
p(\alpha^i \vee \alpha^j \vert \alpha^k I) := p(\alpha^i \vert \alpha^k I) + p(\alpha^j \vert \alpha^k I) - p(\alpha^i \wedge \alpha^j \vert \alpha^k I)
\label{POS}
\end{eqnarray}

If $k \neq i$ and $k \neq j$ then the RHS of (\ref{POS}) is $0+0-0 = 0$.  If $k=i \neq j$ then the RHS of (\ref{POS}) is $1+0-0 = 1$ and if $k=j \neq i$ then it is $0+1-0 = 1$.  If $k=i=j$ then the RHS is $1+1-1 = 1$. This is all as we would expect by intuition.  Similarly,

\begin{eqnarray}
p(\alpha^i \vee \alpha^j \vert \neg \alpha^k I) := p(\alpha^i \vert \neg \alpha^k I) + p(\alpha^j \vert \neg \alpha^k I) - p(\alpha^i \wedge \alpha^j \vert \neg \alpha^k I)
\end{eqnarray}

\noindent is well-defined by construction.  We can also define `$\vee$' relations for the inhomogeneous negations.

\begin{eqnarray}
p(\alpha^i \vee \neg \alpha^j \vert \alpha^k I) &:=& p(\alpha^i \vert \alpha^k I) + p(\neg \alpha^j \vert \alpha^k I) \nonumber \\ & & - p(\alpha^i \wedge \neg \alpha^j \vert \alpha^k I). \\
p(\alpha^i \vee \neg \alpha^j \vert \neg \alpha^k I) &:=& p(\alpha^i \vert \neg \alpha^k I) + p(\neg \alpha^j \vert \neg \alpha^k I) \nonumber \\ & & - p(\alpha^i \wedge \neg \alpha^j \vert \neg \alpha^k I). \\
p(\neg \alpha^i \vee \neg \alpha^j \vert \alpha^k I) &:=& p(\neg \alpha^i \vert \alpha^k I) + p(\neg \alpha^j \vert \alpha^k I) \nonumber \\ & & - p(\neg \alpha^i \wedge \neg \alpha^j \vert \alpha^k I). \\
p(\neg \alpha^i \vee \neg \alpha^j \vert \neg \alpha^k I) &:=& p(\neg \alpha^i \vert \neg \alpha^k I) + p(\neg \alpha^j \vert \neg \alpha^k I) \nonumber \\ & & - p(\neg \alpha^i \wedge \neg \alpha^j \vert \neg \alpha^k I).
\end{eqnarray}

All the above, by construction, give answers consistent with classical probabilistic intuition as long as $d$-realisability and (\ref{PRESERVE2}) are satisfied.  So, if we discuss exhaustive sets of $d$-realisable propositions such that:

\begin{equation}
p(\alpha^i \vert \neg \alpha^j I) = \frac{p(\alpha^i \vert I)}{p(\neg \alpha^j \vert I)} \leq 1 \mbox{ for all } i,j \mbox{ such that } i \neq j,
\end{equation}

\noindent then we can, by construction, get complete consistency with Bayesian reasoning.  History propositions $\alpha^i$ and $\alpha^j$ within such a set are additive over all conditional probabilities even if they are not additive \emph{a priori} such that $p(\alpha^i \vee \alpha^j \vert I) \neq p(\alpha^i \vert I) + p(\alpha^j \vert I)$.  But is there anything wrong with two propositions being additive in one context and not additive in another?  Of course not.  In the Bayesian framework, probabilities are \emph{always} defined contextually \cite{JaynesBOOK} and exclusivity is a contextual property of propositions.

This is not to say that quantum probabilities definitely don't behave in ways that go against classical intuition, only that classical Bayesian probability theory might take us a little further than we may have thought in analysing quantum history propositions.  This approach, which we call Bayesian Histories (BH), has a clear pedagogical basis and, as we shall argue below, may tentatively be experimentally distinguishable from SQT.

So, using BH, we can define history propositions to be exclusive in certain contexts.  But, of course, we can identify these contexts either as \emph{a priori} ones or \emph{a posteriori} ones in reference to Bayes' rule (\ref{BAYES}) depending upon what stage of the Bayesian updating process we are considering.

\section*{A Pedagogical Account of Additivity}

As we discussed above, and recently emphasised by Mana \cite{Mana04}, propositions have certain properties that are contextual.  The exclusivity of propositions is a contextual property of propositions.  Therefore if we have two propositions $A$ and $B$ it is not necessarily the case that they are exclusive in any given context (nor even defined in any given context).  We have defined above that the exclusivity of two propositions arises when $p(A \cap B \vert D) = 0$ and $p(A \cup B \vert D) = p(A \vert D) + p(B \vert D)$ such that all probabilities are well-defined (this happily coincides with the standard Bayesian notion of exclusivity).  In a similar way, consistent historians define `$d$-consistency'; although there is a subtle distinction between the two.  In CH contexts are \emph{defined} to be situations in which $d$-consistency occurs, where-as in BH contexts are far more general.  The exclusivity of propositions might be gained when going from \emph{a priori} probabilities to \emph{a posteriori} probabilities. So, for propositions $A$ and $B$ and prior-information $D$ it might be the case that:

\begin{equation}
p(A \cup B \vert D) \neq p(A \vert D) + p(B \vert D)
\end{equation}

\noindent even though when we update using further information $E$ it is the case that:

\begin{equation}
p(A \cup B \vert E D) = p(A \vert E D) + p(B \vert E D).
\end{equation}

\noindent This is a possibility we can imagine since exclusivity is a contextual property of propositions.  One might be able to define contexts which give additive \emph{a posteriori} probabilities using BH, rather than restricting ourselves to additive \emph{a priori} probabilities (as one might put it when using CH).

Using Bayes' rule we can also na\"{\i}vely derive the following rule:

\begin{eqnarray}
p(A \vert (D \cup E) F) &=& p(D \cup E \vert A F) \frac{p(A \vert F)}{p(D \cup E \vert F)} \\
&=& \frac{p(D \vert A F) p(A \vert F) + p(E \vert A F)p(A \vert F)}{p(D \cup E \vert F)} \label{mum}\\
&=& \frac{p(D \cap A \vert F) + p(E \cap A \vert F)}{p(D \cup E \vert F)}. \label{INVERSE}
\end{eqnarray}

\noindent We get to (\ref{mum}) as long $D$ and $E$ are additive on the \emph{a posteriori} context $A F$.

Throughout the analysis that gave us (\ref{MIXTURE}) we assumed that the context $I$ is well-defined and globally applicable to each null-counterfactual.  This is an assumption that need not be valid.  For example, one could either make the association that $p(\alpha^i \vert C) = d(\alpha^i, \alpha^i)$---identifying the decoherence functional with \emph{a priori} probabilities---or one could associate the decoherence functional with \emph{a posteriori} probabilities:

\begin{equation}
p(\alpha^i \vert (\alpha^k \vee \neg \alpha^k) C) = p(\alpha^i \vert \unit^k C) = d(\alpha^i ,\alpha^i).
\end{equation}

Now, if we associate the decoherence functional with \emph{a posteriori} probabilities then such probabilities are independent of the context $\unit^k$ in which they are taken.  This is a kind of non-contextuality.  Even if the values are the same, however, they may still \emph{behave} differently---probabilities with the same value are not necessarily the same probabilities.  Thus we should keep their notational dependence upon context even if their values are the same.  When can one interpret each $\unit^k$ as a null-counterfactual?  By (\ref{INVERSE}) we have:

\begin{eqnarray}
p(\alpha^i \vert \unit^k C) = \frac{p(\alpha^i \wedge \alpha^k \vert C) + p(\alpha^i \wedge \neg \alpha^k \vert C)}{p(\unit^k \vert C)}.
\end{eqnarray}

So, if we associate the decoherence functional probabilities with \emph{a posteriori} probabilities rather than \emph{a priori} probabilities then we have that:

\begin{equation}
p(\alpha^i \vert \unit^k C) = d(\alpha^i, \alpha^i) = \frac{p(\unit^k \vert \alpha^i C)p(\alpha^i \vert C)}{p(\unit^k \vert C)}.
\end{equation}

This means that $p(\alpha^i \vert \unit^k C) \neq p(\alpha^i \vert C)$.  So, even if probabilities don't depend upon the contexts $\unit^k C$, probabilities still depend on whether such a context is known to be the case or not. \emph{A priori} probabilities are not the same as \emph{a posteriori} probabilities.  It is rather natural to make the association that $I = \unit^k C$ and hence why we must differentiate between $C$ and $I$ in the above presentation.  We reserve the the name `$C$' for \emph{a priori} contexts.  Thus all the na\"{\i}ve probability assignments given in context $I$ can be passed across to probability assignments in contexts $\unit^k C$ for all $k$.

Using (\ref{INVERSE}) we can discuss the probabilities assigned to an exhaustive set of contexts $\vee_k \unit^k$:

\begin{eqnarray}
p(\alpha^i \vert (\vee_k \unit^k) C) &=& \frac{(\sum_k p(\alpha^i \wedge \unit^k \vert C))}{p(\vee_k \unit^k \vert C)} \\ &=& p(\alpha^i \vert \unit^k C) \sum_k p(\unit^k \vert C) = d(\alpha^i, \alpha^i).
\end{eqnarray}

Therefore, a set of contexts $\{\unit^k\}$ that are exhaustive on $C$ gives us the standard probabilities predicted by SQT.

One might now ask how the \emph{a priori} probabilities behave.  We presume that $p(\unit^k \vert \alpha^i C) = 1$ for all $k$ since such conditional probability assignments are natural.  Thus we have that ratios of \emph{a priori} probabilities and ratios of \emph{a posteriori} probabilities are equal, for example:

\begin{equation}
\frac{p(\alpha^i \vert C)}{p(\neg \alpha^k \vert C)} = \frac{p(\alpha^i \vert \unit^k C)}{p(\neg \alpha^k \vert \unit^k C)}. \label{RATIOS}
\end{equation}

In order for the probabilities $p(\alpha^i \wedge \neg \alpha^k \vert C)$ to be well-defined we thus require that such ratios are less than 1.  This is thus equivalent to requiring (\ref{PRESERVE2}).  In order for probabilities $p(\neg \alpha^i \wedge \neg \alpha^k \vert C)$ to be consistent with Bayes' rule we also require that the \emph{a priori} probabilities are quasi-realisable:

\begin{equation}
p(\alpha^i \vert C) + p(\neg \alpha^i \vert C) = L \mbox{ for all } i
\label{QUASIC}
\end{equation}

\noindent where $L$ is a constant.  Since we are assuming that the \emph{a posteriori} probabilities are independent of contexts we require that $p(\alpha^i \vert \unit^k C) = p(\alpha^i \vert \unit^i C)$.  We thus have that $p(\unit^i \vert C) = p(\unit^k \vert C)$. This suggests that all contexts $\unit^k$ are \emph{a priori} equally likely so:

\begin{equation}
p(\unit^i \vert C) = L' \mbox{ for all } i
\end{equation}

\noindent where $L'$ is a constant.  Comparing \emph{a posteriori} and \emph{a priori} probabilities we have that:

\begin{eqnarray}
p(\alpha^i \vert \unit^k C) + p(\neg \alpha^i \vert \unit^k C) = \frac{p(\alpha^i \vert C) + p(\neg \alpha^i \vert C)}{p(\unit^k \vert C)}  = \frac{L}{L'} = K \mbox{ for all } i.
\end{eqnarray}

This is thus completely consistent with our requirement that the \emph{a posteriori} probabilities must be quasi-realisable (\ref{QUASI}).  Thus if $\frac{L}{L'} = K = 1$ then we have $d$-realisable history propositions for all $i$.  If we have $K \neq 1$ then we have a quasi-$d$-realisable set of history propositions.  If $L=L'$ then we have a very cogent interpretation: all the contexts that we invoke consist of $d$-consistent sets and are thus what we have called null-counterfactuals.  So, if $L=L'$, we represent experiments using an equally weighted mixture of null-counterfactuals.  When $K \neq 1$ we don't have a good interpretation so, for now, we reject such cases.

We have a sound interpretation for \emph{a posteriori} contexts when $K=1$, but what does the \emph{a priori} context $C$ refer to?  We don't interpret $C$ here except to say that if Bayesian probability is the correct probability to use then we must require that such \emph{a priori} contexts are consistent with Bayes' rule (\ref{BAYES}).  $C$ is simply some context in which the \emph{a priori} probabilities are well-defined.  $C$ is our knowledge about $\{\unit^k\}$ and our knowledge about $\{\unit^k\}$ is that we don't know which $\unit^k$ happens, so we apply equal \emph{a priori} probabilities.  The standard von Neumann collapse formulation predicts that all probabilities for multi-time measurements are well-defined, but in BH only those that give consistency with Bayesian reasoning are valid.  Thus the collapse hypothesis is not deemed universally valid in BH---it is rather only a convenient hypothesis in certain situations.

Lets look at a standard interference device: a Mach-Zehnder interferometer.  In the standard interpretation there are two possible history propositions which end in detection by a given detector labelled $e$---these histories we call $\alpha^u$ and $\alpha^d$---and SQT predicts that each one happens with probabilities given by the decoherence functional: $d(\alpha^u, \alpha^u)$ and $d(\alpha^d, \alpha^d)$ respectively.  We interpret $\alpha^u$ to be the history proposition that the particle takes the upper path and $\alpha^d$ as the proposition that it takes the lower path.  Thus, in the standard interpretation, the probabilities given by the decoherence functional using these two propositions represent the situation where the path of the particle is measured.  Interference suggests that:

\begin{equation}
d(\alpha^u \vee \alpha^d, \alpha^u \vee \alpha^d) \neq d(\alpha^d, \alpha^d) + d(\alpha^u, \alpha^u).
\end{equation}

\noindent  This means that, in the standard interpretation, when you don't measure the path you predict a different probability at the detector to that you would predict had you measured the path.  One can loosely say then that in one `context' the histories are exclusive and in another they are not, but how do we formalise such notions?  It is clear that in the space of history propositions it is \emph{not} the case that $\neg \alpha^u = \alpha^d$.  We must be more subtle in our use of the negation operation.

Using interpretation \ref{2} we look at this path detection experiment in a subtly distinct fashion.  There are two possible null-counterfactuals $\unit^u = \alpha^u \vee \neg \alpha^u$ and  $\unit^d = \alpha^d \vee \neg \alpha^d$ (lets presume explicitly that $\alpha^u$ and $\alpha^d$ are both $d$-realisable since we have a good interpretation for such propositions). Using (\ref{INVERSE}) we make the association:

\begin{eqnarray}
p(\alpha^u \vert \unit^d C) &=& \frac{p(\alpha^u \wedge \alpha^d \vert C) + p(\alpha^u \wedge \neg \alpha^d \vert C)}{p(\unit^d \vert C)} \\ &=& \frac{0 + p(\alpha^u \vert \neg \alpha^d C)p(\neg \alpha^d \vert C)}{p(\unit^d \vert C)} = \frac{p(\alpha^u \vert C)}{p(\unit^d \vert C)}.
\end{eqnarray}

We can do this as long as the probabilities for $\alpha^d$ and $\neg \alpha^d$ are well-defined and additive on the \emph{a posteriori} context $\alpha^u C$.  With similar provisos we can argue that:

\begin{eqnarray}
p(\alpha^u \vert \unit^u C) &=& \frac{p(\alpha^u \wedge \alpha^u \vert C) + p(\alpha^u \wedge \neg \alpha^u \vert C)}{p(\unit^u \vert C)} \\ &=& \frac{p(\alpha^u \vert C)}{p(\unit^u \vert C)}.
\end{eqnarray}

When we do the experiment we have no \emph{a priori} reason to expect one null-counterfactual to occur over the other so we assign equal weights to each, $p(\unit^u \vert C) = \frac{1}{2} = p(\unit^d \vert C)$.  Each null-counterfactual is deemed to be apt with these \emph{a priori} probabilities.  So the Mach-Zehnder experiment can consist of an equally weighed mixed set $M$ of null-counterfactuals such that:

\begin{eqnarray}
p(\alpha^u \vert M) &=& \frac{p(\alpha^u \wedge \unit^u \vert C) + p(\alpha^u \wedge \unit^d \vert C)}{p(\unit^u \vee \unit^d \vert C)} \\ &=& d(\alpha^u, \alpha^u).
\end{eqnarray}

Thus we recover the SQT predictions for path detection as long as we use $d$-realisable history propositions which give a consistent Bayesian analysis.  Otherwise we must use a different set of null-counterfactuals---the same set of null-counterfactuals can't give use the the case when path detection doesn't occur.  We could also try to define the probability $p(\neg \alpha^u \vert M)$ and in order to do so we would require that the probability $p(\neg \alpha^u \wedge \neg \alpha^d \vert C)$ is well defined, and this requires quasi-realisability.  So, for consistency of the reasoning we use we require at least quasi-realisability for both \emph{a priori} and \emph{a posteriori} probabilities---we require (\ref{QUASIC}) and (\ref{QUASI}) respectively.

We have investigated the situation where the path lengths are equal but, of course, one can easily introduce phase shifters into the arms of the interferometer.  Note that the dynamics is invoked in the very definition of the decoherence functional so phase shifters would be represented by a change in the evolution between to times from standard unitary evolution to one including a change in phase: $d \rightarrow d'$.  Obviously this would have no effect for the path detection experiments but it would have an effect on non-path detection experiments such that $d'(\alpha^u \vee \alpha^d, \alpha^u \vee \alpha^d)$ would depend on a phase factor.  Note $\alpha^u \vee \alpha^d = \alpha^e$ where $\alpha^e$ is the history proposition that the particle is detected by a click in detector $e$ without any path detection.

All this discussion about null-counterfactuals is perhaps rather controversial; it is based around mainly notational issues.  We have invoked them here, however, simply in an attempt to distinguish quasi-realisable and realisable histories.  Even if one does not accept this null-counterfactual formalism we hope that you still take away with you the primary fact that consistency with Bayesian reasoning produces a consistency condition; decoherence probabilities must be at least quasi-realisable and must satisfy (\ref{PRESERVE2}), and $d$-realisable histories seem far less controversial than quasi-$d$-realisable ones.  As to why we should use Bayesian reasoning in the first place, we shall get onto that in a moment once we have discussed linear positivity.

\section*{Quasi-realisability vs Linear Positivity}

Having a rather natural Bayesian interpretation for complete sets of $d$-realisable history propositions, let us now discuss quasi-$d$-realisable history propositions that satisfy (\ref{QUASI}).  These don't give a good interpretation so it is tempting just to reject them, but lets look a little closer at them.  Non-$d$-realisable history propositions simply satisfy the inequality $d(\alpha, \neg \alpha) \neq 0$.  We have that:

\begin{equation}
\re d(\alpha, \neg \alpha) = \re d^{LP}(\alpha) - d(\alpha, \alpha)
\end{equation}

\noindent where $d^{LP}(\alpha)$ is defined on homogeneous history propositions in a similar manner to the decoherence functional (\emph{c.f.} Eq.(\ref{dFUNC})):

\begin{equation}
d^{LP}(\alpha) := \mbox{tr}(C_{\alpha} \rho).
\label{LPFUNC}
\end{equation}

As long as they are positive the $\re d^{LP}(\alpha)$ behave like probabilities.  In the literature they are called Linear Positive (LP) probabilities and were originally promoted by Goldstein and Page \cite{GP95} as a less restrictive alternative to CH probabilities.  Therefore $d$-realisable history propositions have the property that LP probabilities and decoherence functional probabilities have the same value.

Quasi-realisability enforces:

\begin{eqnarray}
p^{LP}(\alpha^i \vert \unit^k C) + p^{LP}(\neg \alpha^i \vert \unit^k C) = K'.
\end{eqnarray}

\noindent Note, however, that LP probabilities are always, by definition, exhaustive when defined on a partition of unity $\sum_i \alpha^i = \unit$ so $K'=1$ for all LP probabilities.  So now we have a choice:  either we attempt to interpret quasi-$d$-realisable propositions or we extend our discussion to LP propositions.  There are a couple of reasons why going the LP way is pedagogically interesting.  Firstly, all LP probabilities are realisable---hence we don't need to worry about non-realisable probabilities cropping up and having to interpret them.  We shall give another reason why we reject non-realisable propositions when we discuss Cox's axioms.  Secondly, LP probabilities are \emph{explicitly} non-contextual; their interpretation doesn't depend upon what other history propositions they are invoked with.  This makes the non-contextuality assumption a bit more explicit such that LP probabilities do not depend upon which null-counterfactuals they are defined with respect to.  So we can, rather naturally, define:

\begin{equation}
p^{LP}(\alpha^i \vert \unit^k C) = p^{LP}(\alpha^i \vert \unit^i C) \mbox{ for all } k,
\end{equation}

\noindent \emph{i.e.} its value is independent of the context, labelled by $k$, we use.  It still depends on the fact that we have a well-defined context hence we keep the notational dependence upon context and don't remove it entirely.  Even if one represents non-contextuality by assuming the contexts are all equivalent and called, say, $I$, then one still gets a consistency condition, namely quasi-realisability, that LP probabilities satisfy since they are realisable.  This non-contextuality assumption is rather analogous to the Gleason non-contextuality (which can also be expressed in terms of null-counterfactuals); afterall, what is non-contextuality if not an assumption that, if you don't know which null-counterfactual you are discussing, that you give each possible null-counterfactual equal \emph{a priori} weighting.  So, there may exist a theorem akin to Gleason's which shows our LP probability assignments to be uniquely defined by certain natural assumptions (although we would have to justify the LP set of history propositions before discussing such a theorem; we do not attempt such a thing here).

So LP probabilities can then be interpreted in a way that is exactly analogous to the way we interpreted the complete sets of $d$-realisable history propositions.  In order for the Bayesian probability assignments to be well-defined probabilities bounded by $0$ and $1$ then we must, in analogy with (\ref{PRESERVE2}), require that all LP probabilities preserve the partial order on the history space for all LP history propositions:

\begin{equation}
\alpha^i \leq \neg \alpha^j \Rightarrow \re d^{LP}(\alpha^i) \leq \re d^{LP}(\neg \alpha^j) \mbox{ for all } i \neq j.
\label{PRESERVE3}
\end{equation}

\noindent This is satisfied for all LP history propositions.

We can define $K' = \frac{L''}{L'''}$ in an analogous way such that:

\begin{equation}
p^{LP}(\alpha^i \vert C) + p^{LP}(\neg \alpha^i \vert C) = L'' \mbox{ for all } i
\end{equation}

\noindent and

\begin{equation}
p^{LP}(\unit^k \vert C) = L''' \mbox{ for all } k.
\end{equation}

For LP history propositions we have that $K'=1$ and thus that $L''=L'''$.  Thus we interpret the \emph{a priori} context $C$ to be the knowledge that we have no knowledge about the contexts $\unit^k$ and thus assign them equal \emph{a priori} probabilities.  Thus we can, if we wish, extend BH to include all LP history propositions and not just $d$-realisable propositions (which are, of course, also LP).

If BH is correct then it helps, in part, to `explain' interference because the probabilities invoked obey rules that are consistent with our classical intuition.  If BH is incorrect---if non-LP history propositions remain well-defined and experimentally realisable---then we have a theory that obeys our classical intuition, to some extent at least, which SQT disobeys---this in itself would be a novel result.

\section*{Why Bayes' Rule?}

Having shown that there is a certain amount of consistency between Bayes' rule and the LP formalism the following programme presents itself: perhaps we can derive the LP probabilities by taking the history algebra and applying something akin to Cox's axioms \cite{CoxBOOK} to this space.  Cox's work derives probability theory over an underlying Boolean algebra using simple consistency conditions that a natural form of inductive reasoning should obey, so it seems natural just to try and apply a similar kind of reasoning with the history algebra.  If such a proof is found and as long as the history algebra could then be justified by \emph{a priori} means---or by simple physically justified axioms---then one would be able to prove that the LP formalism is \emph{just} another kind of probability theory.

It is clear, however, that our na\"{\i}ve assumption of using Bayes' rule is justified by Cox's axioms.  Cox's first axiom is that the probability of a statement conditional upon some hypothesis determines the negation of that same statement upon the same hypothesis.  The second axiom, more relvent here, is that the probability that two statements are both true upon a given hypothesis is determined alone from the probability of one of the statements conditional upon the given hypothesis and the probability of the other statement conditional upon the hypothesis conjoined with the presumption that the first statement is true.  In our notation this is written schematically as:

\begin{equation}
p(\alpha \wedge \beta \vert I) := F[p(\beta \vert I), p(\alpha \vert \beta I)]
\label{Cox2}
\end{equation}

\noindent where $F$ is an arbitrary function to be determined that is sufficiently well-behaved for our purposes.

The underlying algebra for history propositions is associative so the following statement is true:
\begin{equation}
\alpha \wedge (\beta \wedge \gamma) = (\alpha \wedge \beta) \wedge \gamma = \alpha \wedge \beta \wedge \gamma.
\label{property}
\end{equation}

The above property (\ref{property}) forces $F$ not to be arbitrary and Cox \cite{CoxBOOK} proves that Bayes' rule is a consequence (Jaynes highlights a more general proof in \cite{JaynesBOOK}):

\begin{equation}
p(\alpha \vert \beta I) = \frac{p(\alpha \wedge \beta \vert I)}{p(\beta \vert I)}.
\label{properbayes}
\end{equation}

Since the associativity of `$\wedge$' (\ref{property}) is valid for our quantum logics then Bayes' rule follows and the above work is justified to an extent.  Note that for homogeneous histories defined over the same temporal support we have that $\alpha \wedge \beta = \beta \wedge \alpha$ so the histories equivalent of the multiplication rule (\ref{AND}) also follows in such cases.

Thus, although the above analysis initially seems quite na\"{\i}ve there is some truth to it---Bayes' rule, if nothing else, should be obeyed by any natural notion of probability by Cox's axioms.  We have, however, yet to generalise Cox's other proofs to the HPO algebra of history propositions proper.  This remains work in progress.  It is clear that the decoherence probabilities (which are also the standard probability assignments invoked using von Neumann collapse), at least in the Hamiltonian formulation, need not always obey Bayes' rule---they can disobey (\ref{CONDITION}) for example and need not be realisable or quasi-realisable---so we have to restrict our attention to either $d$-realisable histories or LP ones (or use some other assignment).

A na\"{\i}ve application of Cox's first axiom,

\begin{equation}
p(\neg \alpha \vert I) := G[p(\alpha \vert I)],
\label{Cox1}
\end{equation}

\noindent
suggests we should use a fixed value of $K$, hence why we have restricted our attention to realisable histories.  Hence we should use realisable probabilities that obey Bayes' rule---we should use LP probabilities.  This approach may not be considered wholly satisfactory because we are not giving probability assignments to all histories but only the LP subset of the algebra.  This is curiously analogous to the situation in Youssef's work \cite{Youssef01} where, in deriving a form of SQT as an `exotic' complex probability theory, he has to presume a subset where the standard real probabilities are manifested.  We have placed the term `exotic' in scare quotes because we prefer not to use the term.  When invoking Bayesian reasoning there is no \emph{a priori} reason probabilities should be real numbers (also see \cite{Isham02}).  We only need to presume they are real when notions of relative frequency are applicable.  Hence we would rather call such theories just probability theories;  there is nothing really `exotic' about them.  Hence we leave open the possibility of deriving the whole of the histories formalism in such a manner.  We investigate such a possibility in forthcoming work.

\section*{Experimental Differentiation}

So if we \emph{re-define} multi-time measurements to be equally weighted sets of null-counterfactuals (due to some principle of insufficient reason) we can get all Linearly Positive (LP) probabilities.  One might wish to take this very seriously.  There are two tacks that we can take in regards to BH.  Firstly, we could choose to use BH to discuss closed quantum systems.  LP probabilities were originally promoted in this manner \cite{GP95} because as soon as we discuss closed quantum systems then using Eq.(\ref{dFUNC}) to assign probabilities to history propositions simply becomes a \emph{postulate}. Using the real part of Eq.(\ref{LPFUNC}) to assign candidate probabilities is another, equally valid, postulate.  And, of course, any rule that is distinct from the von Neumann projection postulate must be investigated very carefully. Di\'{o}si \cite{Diosi94} has argued that the LP probabilities should not be used as probability assignments because they are not consistent with the statistical independence of subsystems.  Implicit in this critique is the use of a relative frequency interpretation of probability but it is not clear that using a relative frequency interpretation for closed systems is wholly sound.  The only other option we have (propensities being simply objectivised relative frequencies) is to use Bayesian probabilities; and if we do use Bayesian probabilities then LP probabilities are \emph{promoted} over decoherence probabilities as they have a very simply interpretation and obey Cox's axioms for the LP subset.  Bayesian probability theory encompasses the use of relative frequencies in certain situations \cite{JaynesBOOK} so there is nothing necessarily untoward about this.

Secondly, we could try and apply BH to actual experiments.  Anastopoulos suggests, in \cite{Anast04}, that we should try and experimentally check that $d$-inconsistent sets do really make good statistical sense.  With a similar emphasis it may also be prudent just to check whether non-LP history propositions do really make good statistical sense in quantum experiments.  But, of course, what do we mean by ``good statistical sense''?  If we assume that ``good statistical sense'' is equivalent with the statement ``is consistent with Bayes' rule'' then BH is promoted as a tautology.  Otherwise one must use a form of statistics that is inconsistent with Bayes' rule and is also well-defined.

So, it is clear that, at present, the only way we know how to get relative frequencies out from quantum history theories is by discussing $d$-consistent sets.  Those sets that aren't $d$-consistent may not give well-defined notions of relative frequency \cite{Anast04}.  This might be because such relative frequencies aren't convergent or converge to many different values \cite{Aerts02}.  Of course, if relative frequencies converge to many values then the most natural interpretation is to suggest that we are inadvertently or necessarily mixing contexts.  Hence we should, as argued above, be very careful about the notation we use and include any contextual dependence in the very definition of the probabilities involved.

Hartle \cite{Hartle04}, for example, has recently analysed the double slit experiment in reference to LP history propositions and shows that if the resolution of the screen is sufficiently high then the candidate probabilities predicted using the real part of Eq.(\ref{LPFUNC}) will not be well-defined.  But resolution coarser than a critical value will give well-defined LP probabilities. How seriously should we take this?  Normally such probabilities are interpreted in closed systems but should we not just check that these aren't compatible with the relative frequencies of actual experiments?  We have argued that LP probabilities can be interpreted in a particularly Bayesian way; the next challenge for BH is thus to try and work out how such Bayesian probabilities are related (obviously such a relation might be non-trivial) to the relative frequencies of experiments---as this might provide a way to experimentally distinguish the two approaches.  There are a variety of ways one can derive relative frequencies from Bayesian probabilities; for example one can invoke notions of exchangability, independence or use maximum entropy methods \cite{JaynesBOOK}.  Statistical independence in history theories has been studied by Di\'{o}si \cite{Diosi94} and discussed by Hartle \cite{Hartle04} but there may be other useful ways to invoke relative frequencies from Bayesian probabilities.

To the present author, it is tempting to believe in BH simply for the cogency of the interpretation.  It uses standard notions of Bayesian probability that are well understood and it pedagogically invokes the contextuality implicit in the propositional nature of history propositions.  Although, of course, further investigation and statistical analysis of experiments are necessary to justify it above the standard interpretation.  In the standard interpretation \emph{any} ordered set of single-time measurements is realisable regardless of problems of non-additivity (presuming the relevant apparatus can be made).  In BH, only those multi-time measurements that give well-defined \emph{a posteriori} and \emph{a priori} probabilities are experimentally realisable with good statistics.  Thus the standard interpretation and BH give distinct statistical predictions when interpreted instrumentally.  But, of course, the instrumental validity of BH bares little relation to whether BH should be invoked when discussing closed quantum systems.  In closed systems probability is implicitly used as a form of inference rather than as relative frequencies of experiments so one should naturally use Bayesian probability.  We have shown that all LP probabilities are consistent with Bayesian reasoning, whereas not all probabilities of the form (\ref{Trace}) are (when using the natural space of history propositions).

\section*{Entropy}

By invoking contexts in which history propositions are exclusive and exhaustive we now have the opportunity to use standard Shannon entropy to quantify information.  For example, if a set $\{\alpha^i : i=1,2,...,N_\alpha\}$ is exclusive and exhaustive on \emph{a priori} context $C$ then we can define the the set's Shannon entropy in a simple way.  Here we denote probabilities with a small $p$ and probability distributions with a large $P$.  The Shannon entropy is then simply given by:

\begin{equation}
H[P(\alpha^i \vert C)] := -K_H \sum_{i=1}^{N_\alpha} p(\alpha^i \vert C) \ln p(\alpha^i \vert C).
\end{equation}

\noindent where $K_H$ is a constant.  We cannot define such an entropy for sets $\{\alpha^i\}$ that aren't exhaustive and exclusive on $C$.  But we can define an entropy for them if we take an \emph{a posteriori} context $I$ in which $\{\alpha^i\}$ \emph{are} exclusive and exhaustive:

\begin{equation}
H[P(\alpha^i \vert I)] = -K_H \sum_{i=1}^{N_\alpha} p(\alpha^i \vert I) \ln p(\alpha^i \vert I).
\end{equation}

In \cite{Mana04}, Mana has cogently argued against improper use of such entropy concepts in SQT.  Since we have used standard Bayesian probability and kept contextuality in check we can use Mana's pedagogical results in the histories domain as well.  As such, if we have two sets of propositions $\{\alpha^i\}$ and $\{\beta^j\}$ that are exclusive and exhaustive in the \emph{same} context $I$---they are both `sets of alternatives' in $I$---then we can define the conditional entropy as follows:

\begin{eqnarray}
H[P(\beta^j \vert \alpha^i I)] &:=& -K_H \sum_{i=1}^{N_\alpha} p(\alpha^i \vert I) H[P(\beta^j \vert \alpha^i I)] \\ &=& -K_H \sum_{i=1}^{N_\alpha} p(\alpha^i \vert I) \sum_{j=1}^{N_\beta} p(\beta^j \vert \alpha^i I) \ln p(\beta^j \vert \alpha^i I).
\end{eqnarray}

An analogous definition is used for $H[P(\alpha^i \vert \beta^j I)]$. In such a case the following standard formulae should apply by mathematical necessity \cite{Mana04}:

\begin{eqnarray}
H[P(\alpha^i \wedge \beta^j \vert I)] &=& H[P(\alpha^i \vert I)] + H[P(\beta^j \vert \alpha^i I)] \\ &=& H[P(\beta^j \vert I)] + H[P(\alpha^i \vert \beta^j I)]
\end{eqnarray}
\begin{equation}
H[P(\beta^j \vert I)] \geq H[P(\beta^j \vert \alpha^i I)].
\end{equation}

These are the standard strong additivity and concavity properties of Shannon entropy.  We can avoid any of the confusions highlighted by Mana \cite{Mana04} by using such standard definitions of Shannon entropy.

If we interpret multi-time measurements as equally weighed mixed sets of null-counterfactuals then such entropy concepts allow us to compare $d$-consistent or LP sets entropically.  This is not particularly useful if one interprets history propositions in the standard quantum cosmological manner but, of course, it may be very useful when interpreting quantum history propositions instrumentally.  The reader is also referred, in earnest, to a Bayesian derivation of entropic concepts given recently by Caticha \cite{Catich03}.

\section*{Future Research}

Isham's seminal work on CH and topos theory \cite{Isham97} pre-empts the idea that $d$-inconsistent sets can be assigned a certain amount of meaning using a notion of $d$-accessability which is related to our definition of $d$-realisability\footnote{Note that our use of the term `$d$-realisable' in not the same as its use in \cite{Isham97}.}.  The present work can be considered a pedagogical account of such toposophic concepts in the domain of instrumental quantum theory, which shows that such a generalisation provides different statistical predictions.  Isham also argued that it is pedagogically useful to discuss $d$-consistent Boolean algebras rather than $d$-consistent sets \emph{per se} because such objects are more akin to what we think of in classical probability theory.  We agree with this sentiment (although we didn't submit to it here because of the useful illustrative notion of null-counterfactual) and the above work can easily be framed as such: one can define Boolean sub-algebras $W$ that consist of history propositions; $W = \{\alpha^i : i=1,2,....,M\}$ to be $d$-consistent if \cite{Isham97}:

\begin{equation}
d(\alpha^i \wedge \alpha^j, \alpha^i \wedge \alpha^j) = d(\alpha^i, \alpha^j) \mbox{ for all } \alpha^i,\alpha^j \in W.
\end{equation}

\noindent Furthermore, in our notation, we can ask that:

\begin{equation}
p(\alpha^i \vee \alpha^j \vert I) = p(\alpha^i \vert I) + p(\alpha^j \vert I) - p(\alpha^i \wedge \alpha^j \vert I) \mbox{ for all } \alpha^i,\alpha^j \in W.
\label{61}
\end{equation}

The extended definition of entropy for not-necessarily exclusive events given by Cox \cite{CoxBOOK} can also be applied to such propositions in a Boolean algebra as long as (\ref{61}) is satisfied---when such propositions are not-exclusive they would be not-exclusive in the same way that classical propositions can be not-exclusive.  It might also be pedagogically useful to generalise single-time von Neumann null-counterfactuals to Boolean algebras proper.  One can discuss more general contexts in which $d$-inconsistent Boolean algebras are consistent---in an \emph{a posteriori} sense---with rules (\ref{BAYES}) and (\ref{AND}).  The present author is not yet sure exactly how such toposophic concepts are related to BH; this is left for further research.  Nor is it clear how such concepts pass across to LP history propositions.

Operational notions \cite{KrausBOOK, DaviesBOOK} such as Positive Operator Valued Measures (POVMs) provide a generalisation of von Neumann single-time measurements in the sense that each POVM defines a set of propositions that are apt in a measurement with certain probabilities.  As such, it is easy to imagine a operational generalisation of the above work (see \cite{Rudolph96, Kent98}).  In the POVM formalism single-time propositions are represented by effect operators that need not necessarily be orthogonal.  When discussing such operational notions it is important to distinguish between `orthogonal' propositions and `exclusive' ones---POVMs can consist of non-orthogonal propositions but these propositions are interpreted to occur exclusively regardless of whether they are orthogonal or not.  In SQT we can prepare a mixed state of non-orthogonal pure states such that each pure state occurs exclusively in the mixed state with a given weight; similarly POVMs can consist of exclusive non-orthogonal propositions.  So, if we interpret the outcomes of POVMs to happen exclusively, a generalisation into the multi-time domain that is compatible with interpretation \ref{2} might be possible.  Such a multi-time generalisation, however, would require a logic to the set of effect operators akin to the quantum logic of projection operators.  Time evolution in the POVM formalism is more general than unitary evolution which might add an extra complication.

One is also tempted to apply such null-counterfactuals to Bell-like experiments.  It is exactly a cogent notion of a null-counterfactuals that is lacking in such analyses \cite{Shimon04}.  By describing such experiments in terms of realisable sets of history propositions one might nullify any proofs of nonlocality.  We have shown above that there is no \emph{a priori} reason why all multi-time candidate propositions made out of single-time propositions should be well-defined consistently.  Similarly, there is no \emph{a priori} reason why candidate null-counterfactual propositions about multiple spacelike separated spacetime regions must all be well-defined (as is implicitly assumed in \cite{Stapp03} and criticised by \cite{Shimon04}).  By using interpretation \ref{2}, null-counterfactual statements might necessarily be statements about both spacelike separated regions and cannot be well-defined for individual small spacetime regions.  This would, tentatively, be a way to argue against the EPR paper \cite{EPR} in a way akin to Bohr's response \cite{Bohr35}.  It may also be a way to promote Bayesian probability over relative frequencies \cite{Marlow05b,Marlow05}.  This is presently left for further research; as are relativistic generalisations of BH.

\section*{Conclusions}

We have shown that the two interpretations of multi-time measurements given by Anastopoulos \cite{Anast04} can be distinguished by how they treat non-realisable history propositions.  If we assume that multi-time measurements consist of successions of single-time measurements then one gets non-additive (and thus non-exclusive) propositions---this is the standard interpretation of multi-time measurements.  Alternatively, if we assume that multi-time measurements are made up of sets of exclusive and exhaustive history propositions (and recover single-time SQT when using single-time history propositions) then one promotes a more standard notion of probabilistic exclusivity. The latter interpretation seems cogent and it might be experimentally differentiated from the former by a statistical analysis of non-realisable propositions in experiments.  If the probabilities of non-realisable propositions all remain well-defined then we must stick to the standard interpretation, but otherwise the latter novel interpretation would be promoted.  Since the latter interpretation provides a certain amount of philosophical clarity over the former, it is worthwhile trying to experimentally distinguish the two.  We justify our novel approach, in part, by invoking Cox's probability axioms on the history algebra and showing that Bayes' rule should be obeyed by any natural probability assignments.

\section*{Acknowledgements}

I would like to thank George Jaroszkiewicz for his kind patience, EPSRC for funding this work, and an anonymous referee for useful comments.

All preprints refer to the http://arxiv.org website.

\end{document}